\newif\ifdraft
\drafttrue
\draftfalse

\ifdraft
\newcommand{\lenSect}[1]{{\color{blue}[#1 Seite(n)]}}
\documentclass{llncs}
\usepackage{a4wide}
\else
\newcommand{\lenSect}[1]{}
\documentclass{llncs}
\fi

\renewcommand{\qed}{\hfill \ensuremath{\Box}}
\let\oldendproof\endproof
\def\endproof{\qed\oldendproof}

\usepackage{xcolor} 
\usepackage[ascii]{inputenc}
\usepackage{amsmath,amssymb} 
\usepackage{enumerate}
\usepackage{todonotes}
\usepackage{multirow,tabularx}

\usepackage{url}
\usepackage{algorithm2e}

\newcommand{\fworkName}{PriCL}
\renewcommand\paragraph[1]{\smallskip\noindent\textbf{#1.}}

\title{\fworkName: Creating a Precedent\\ A Framework for Reasoning about
  Privacy Case Law}

\author{Michael Backes \and Fabian Bendun \and J\"org Hoffmann \and
  Ninja Marnau}

\institute{CISPA, Saarland University\\
  \email{\{backes,bendun,hoffmann,marnau\}@cs.uni-saarland.de}}

\newif\iflongversion
\longversiontrue

\newcommand{\cutShort}[1]{%
\iflongversion%
#1%
\fi}

\newcommand{\replShort}[2]{%
\iflongversion%
#1%
\else%
#2%
\fi}

\newcommand{\AND}{\textsf{AND}}
\newcommand{\OR}{\textsf{OR}}

\newcommand{\mayref}{\textsf{may-ref}}
\newcommand{\mustref}{\textsf{must-agree}}

\newcommand{\ALC}{\mathcal{ALC}}
\newcommand{\exptime}{\text{ExpTime}}

\newcommand{\confPred}{\textsf{df}}
\newcommand{\DB}{\textsf{DB}}

\newcommand{\caseDesc}{\textsf{CaseDesc}}
\newcommand{\prTree}{\textsf{ProofTree}}

\newcommand{\labAxiom}{\textsf{Axiom}}
\newcommand{\labDec}{\textsf{Assess}}
\newcommand{\labCase}{\textsf{Ref}}

\newcommand{\sub}{\textsf{sub}}
\newcommand{\subcase}{\textsf{sub}}
\newcommand{\legalPred}{\textsf{is\_legal\_action}}

\newcommand{\refer}{\textsf{references}}
\newcommand{\KB}{\textsf{KB}}
\newcommand{\courtSet}{\textsf{Courts}}
\newcommand{\crt}{\textsf{crt}}
\newcommand{\pre}{\textsf{pre}}
\newcommand{\fact}{\textsf{fact}}
\newcommand{\pres}{\textsf{pres}}
\newcommand{\facts}{\textsf{facts}}

\newcommand{\concl}{\textsf{fact}}

\newcommand{\actionSet}{\textsf{Actions}}

\newcommand{\leref}{<_{\textsf{ref}}}
\newcommand{\lecourt}{<_{\S}}

\newcommand{\abs}[1]{\vert#1\vert}

\newcommand{\NP}{{\bf{NP}}}
\newcommand{\coNP}{\bf{coNP}}

\newcommand{\phipre}{\phi^{\textsf{preSwitch}}}
\newcommand{\phifact}{\phi^{\textsf{factSwitch}}}
\newcommand{\phia}{\phi^{\textsf{(1)}}}
\newcommand{\phib}{\phi^{\textsf{(2)}}}
\newcommand{\phic}{\phi^{\textsf{(3)}}}
\newcommand{\chosen}{\textsf{chosen}}
\newcommand{\psic}{\psi^{(3)}}

\newcommand{\conFills}{\textbf{fills}}
\newcommand{\conOneOf}{\textbf{one-of}}
\newcommand{\roleAnd}{\textbf{role-and}}
\newcommand{\roleNot}{\textbf{role-not}}
\newcommand{\roleProd}{\textbf{product}}
\newcommand{\roleInverse}{\textbf{inverse}}

\begin{document}

\maketitle



\begin{sloppypar}
\begin{abstract}
  We introduce \fworkName: the first framework for expressing and
  automatically reasoning about privacy case law by means of
  precedent. \fworkName\ is parametric in an underlying logic for
  expressing world properties, and provides support for court decisions,
  their justification, the circumstances in which the justification
  applies as well as court hierarchies. Moreover, the framework offers
  a tight connection between privacy case law and the notion of norms
  that underlies existing rule-based privacy research.  In terms of
  automation, we identify the major reasoning tasks for privacy cases
  such as deducing legal permissions or extracting norms.  For solving
  these tasks, we provide generic algorithms that have particularly
  efficient realizations within an expressive underlying logic.
  Finally, we derive a definition of deducibility based on legal
  concepts and subsequently propose an equivalent characterization in
  terms of logic satisfiability.

\end{abstract}

\section{Introduction}

Privacy regulations such as HIPAA, COPPA, or GLBA in the United States
impose legal grounds for privacy~\cite{hipaa,glba,coppa}. In order to
effectively reason about such regulations, e.g., for checking
compliance, it is instrumental to come up with suitable formalizations
of such frameworks along with the corresponding automated reasoning
tasks.

There are currently two orthogonal approaches to how regulations are
expressed and interpreted in real life that both call for such a
formalization and corresponding reasoning support.
One approach is based on \cutShort{providing an} explicit
\cutShort{set of} rules that define what is allowed and what is
forbidden. The alternative is to consider precedents\cutShort{ and
  case law}, which is the approach predominantly followed in many
countries such as the US.
Precedents are cases that decide a specific legal context for the
first time and thus serve as a point of reference whenever a future
similar case needs to be decided. Moreover, even judges in countries
that do not base their legal system on precedents often use this
mechanism to validate their decision or shorten the process of
argumentation.

Case law is particularly suitable for resolving vague formulations
that naturally occur in privacy regulations like the definition of
`disclosure' in COPPA\cutShort{: ``The term `disclosure' means [...]
  the release of personal information collected from a child in
  identifiable form''}. Here, case law could reference decisions that
define what circumstances are qualified as a non-identifiable form of
personal data, thereby aiding the
user 
by providing judicially accurate interpretation of such terms.

While rule-based frameworks have received tremendous attention in
previous research (see the section on related work below) there is currently no formalization for case law that is amenable to automated
reasoning.

\paragraph{Our contribution} Our contribution to this problem space is
threefold:
\vspace{-\topsep}\begin{itemize}
\item We derive important legal
  concepts from actual judicial processes and relevant
  requirements from related work. The resulting framework \fworkName,\  can be applied to the judicature of many different countries as it does not
  assume any specific argumentation.
\item We tailor the framework for privacy regulations. In
  particular, our privacy specific case law framework is compatible
  with former policy languages since it has only minimal requirements
  regarding the logic. Therefore, it is possible to embed other
  formalizations into our framework.
\item We define the major reasoning tasks that are needed to
  apply the framework to privacy cases. In particular, these tasks
  allow us to derive requirements for the underlying logic which we
  analyze. Several logics allow an embedding of the reasoning tasks by
  giving an equivalent characterization of the tasks. Consequently, we
  are able to select a well suited logic.
\end{itemize}\vspace{-.5\topsep}

\cutShort{\noindent In total, the case law framework that we introduce gives a
new approach for compliance with privacy regulations. In particular,
it makes it possible to implicitly use any regulation if it was
previously referenced by a judge. Moreover, it also provides for
reasoning tasks in cases where no regulation is applicable but
judicial precedents exist.}

\paragraph{Related work}
There are plenty of privacy regulations that companies are required to
comply with. In the US there are regulations for specific sectors,
e.g., HIPAA for health data, COPPA for children's data, or GLBA and
RFPA for financial data. In the EU, the member states have general
data protection codes. The legislative efforts to harmonize these
national codes via the EU Data Protection Regulation~\cite{eudatareg}
are proceeding and already provide for identifying legislative
trends. The importance and impact of these privacy regulations has
brought the interpretation thereof to the attention of more
technically focused privacy
research~\cite{lammel2013understanding,barth2006privacy,%
  annas2003hipaa,duma2007privacy,breaux2008analyzing,oh2014privacy}.

Policy languages were mainly developed in order to model these
regulations and to reflect companies' policies. Many of the modern
logics modeling regulations are based on temporal
logic~\cite{garg2011policy,basin2013monitoring,datta2011understanding,tschantz2012formalizing,barth2007privacy}
and were successfully used to model HIPAA and
GLBA~\cite{deyoung2010logical}\cutShort{ and should be applicable to other
regulations as well}. While these logics focus on expressiveness in
order to reflect the regulations, the logics for company policies
focus on enforcement~\cite{backes2004efficient,ashley2003enterprise}
and thus also on
authorization~\cite{anderson2005comparison,ashley2003enterprise}. Consequently,
company policies are mostly based on access control
policies~\cite{ni2010privacy,karat2009policy}.

Bridging the gap between the regulation policies and the company's
policies leads to automating compliance
checks~\cite{senbootstrapping}. For many deployed policies, i.e., the
ones that are efficiently enforceable, this is currently not possible
due to the lack of decidability regarding the logics used to formalize
regulations. However, for these cases there exist run-time monitoring
tools that allow compliance auditing on log
files~\cite{barth2006privacy,garg2011policy,basin2008runtime,basin2013monitoring}. In
particular, such auditing was invented for
HIPAA~\cite{garg2011policy}.

A different approach for achieving compliance is guaranteeing
privacy-by-design~\cite{maffei2013security,cavoukian2012privacy,gurses2011engineering}. However,
the policy of these systems still needs to be checked for compliance
with the relevant privacy regulations.

\cutShort{There is also an orthogonal approach when designing privacy policies
that focus on the end user, i.e., designing a policy that is formal
and can be formulated in an user-understandable
way~\cite{AnBerLiYu:07:roadmap}. First attempts using
P3P~\cite{cranor2003p3p,salim2007enforcing,ashley2004enforcement} were
unsuccessful. However, it is important to incorporate the user in the
process of policy design in order to gain her
trust~\cite{kelley2009nutrition,flavian2006consumer}.}



\section{Ingredients\lenSect{1}}\label{sec:ingredients}

In the first step we illustrate which components are essential for a
case law framework. To that end, we analyze actual judicial processes
and derive ingredients for the framework from the relevant legal
principles. \cutShort{In particular, the court decision and its
  justification give insights into how the decision is made and which
  judicial concepts have to be reflected by our framework.} Hence, in
the following, we analyze a representative court decision\footnote{The
  quotes are taken from MARTINO v. BARNETT, Supreme Court of Appeals
  of West Virginia, No.~31270, Decided: March 15, 2004. The decision
  text is public at
  \url{http://caselaw.findlaw.com/wv-supreme-court-of-appeals/1016919.html}.}
and discuss the implications for our framework.

\paragraph{The conflict} {\it ``This matter involves three certified
  questions from the Circuit Court of Harrison County regarding
  whether applicable state and federal privacy laws allow
  dissemination of confidential customer information \replShort{by
    an insurance company to an unaffiliated third party}{[...]} during
  the adjustment or litigation of an insurance claim.''}

Every case reaching a court is based on a conflict, i.e., there is
some question, as the one above, for which different parties have
different opinions on its truth value.\cutShort{\footnote{In the example case,
  the parties are a plaintiff, who was injured in a car accident, and
  an insurance company, which refused to disclose the home address of
  the other person involved in the accident. The insurance company
  claimed that to do so would violate the privacy provisions of the
  Gramm-Leach-Bliley-Act (GLBA) and the West Virginia Insurance
  Commission's Privacy Rule.}} As a requirement for the framework, we
can conclude that there has to be a conflict that needs to be resolved
by a decision. This decision can be an arbitrary statement; hence, we
call it a \emph{decision formula}.

\paragraph{Sub-cases} 
A decision's justification usually involves decisions of several {\it
  sub-cases} in order to arrive at the final decision formula,
e.g. the court needs to decide whether a specific law is applicable
before examining what follows from its application. Each of these
individual sub-case decisions may become a precedent for decisions
which deal with a similar sub-case.

\paragraph{The circumstances} {\it ``[The plaintiff] concedes that
  under the definitions of the GLBA [...] information he requests is
  technically nonpublic personal information of a customer which the
  Act generally protects from disclosure\replShort{ to nonaffiliated third
  parties}{[...]}.''}

Every case contains some factual background. These facts constitute
some statements which are not under discussion but measurably true,
e.g., that an address is nonpublic personal information. We summarize
these facts in a \emph{case description}.

\paragraph{Referencing related court decisions} {\it ``[T]he United
  States District Court for the Southern District of West Virginia
  handed down an opinion in Marks v. Global Mortgage Group, Inc., 218
  F.R.D. 492 (S.D.W.Va.2003), providing us with timely and pertinent
  considerations.''}

The key of case law is referencing other cases in order to derive
statements. In the example case, this capability is used to introduce
an argumentation from a different court. This mechanism is also used
when statements are derived from
regulations. 
%
Consequently, the framework has to be capable of introducing
statements during the case justification by \emph{references} to their
origin.

\paragraph{Argumentation structure of the justification} {\it ``[The]
  GLBA provides exceptions to its notification and opt-out procedures,
  including [...]''}

The argumentation structure of the justification is not linear, i.e.,
of the form $A\Rightarrow B\Rightarrow\ldots\Rightarrow$. But the
arguments can be ordered in a tree form. The exceptions stipulated by
the GLBA are enumerated and then discussed in the case
justification. If more than one is applicable, these may serve as {\it
  independent decision grounds}, each being a potential precedent in
its own right.\cutShort{\footnote{O'Gilvie v. United States, 519 U.S. 79, 84
  (1996).} As a consequence, we believe that a \emph{proof
    tree} fits the overall structure best.}
 
\paragraph{World knowledge} {\it ``[We] conclude that nonpublic
  personal information may be subject to release pursuant to judicial
  process.''}

In the argumentation, the court leaves to the reader's knowledge that
the plaintiff's litigation actually is a ``judicial process''. These
open ends in the argumentation are neither explicitly covered by a
decision nor by a case reference. Therefore, we need some world
knowledge $\KB_W$ that will cover these axiomatic parts of the
argumentation.

\paragraph{Precedents and stare decisis}
The doctrine of {\it stare decisis} (to stand by things decided) or
binding precedents is unique to common law systems. The decisions of
superior courts are binding for later decisions of inferior courts
({\it vertical stare decisis}). These binding precedents are applied
to similar cases by analogy.

\cutShort{A special case is the binding nature of previous decisions
  on the same hierarchical level or by the deciding court itself ({\it
    horizontal stare decisis}). While the details of binding
  precedents of different courts on the same level is subject to an
  ongoing scholarly debate, a {\it court reversing itself} is a more
  infrequent occurrence but usually has high impact (for example, in
  the years 1946-1992, the U.S. Supreme Court reversed itself in 130
  cases\footnote{Congressional Research Service --- Supreme Court
    Decisions Overruled by Subsequent Decision
    (1992). \url{http://www.gpo.gov/fdsys/pkg/GPO-CONAN-1992/html/GPO-CONAN-1992-13.htm}
    The U.S. Supreme Court has explained its practice as follows:
    ``[W]hen convinced of former error, this Court has never felt
    constrained to follow precedent.'' --- Smith v. Allwright, 321
    U.S. 649, 665 (1944)}) and needs to be reflected in our
  framework.\footnote{Federal and state supreme courts are allowed to
    overrule their own precedents. State Oil Co. v. Khan, 522 U.S. 3,
    20 (1997); Freeman \& Mills, Inc. v. Belcher Oil Co., 11 Cal. 4th
    85, 93 (1995).}}

In addition to the binding precedent, there also exists the persuasive
precedent:
{\it ``While we recognize that the decision of the Marks court does
  not bind us, we find the reasoning in Marks regarding a judicial
  process exception to the GLBA very persuasive and compelling''.}

\cutShort{Here, a court is not bound by an earlier decision, in our example because
the earlier decision was made by an inferior court, but finds the
argumentation so persuasive that it is voluntarily used as a
precedent.}

Stare decisis does not apply in civil law systems, like those of
Germany or France. However, these systems have a {\it jurisprudence
  constante}, facilitating predictable and cohesive court
decisions. Though civil law judges are not obliged to follow
precedents, they may use prior decisions as persuasive precedents and
oftentimes do so.

\paragraph{Material difference}
Stare decisis only applies if the subsequent court has to decide on a
case or sub-case that is similar to the precedent. Therefore, if the
court finds {\it material difference} between the cases, it is not
bound by stare decisis. In practice, judges may claim material
difference on unwarranted grounds, which may lead to conflicting
decisions of analoguous cases within our framework. Thus, we need to
be able to account for {\it false material difference}.

\paragraph{Involving court hierarchies} {\it ``[W]e look initially to
  federal decisions interpreting the relevant provisions of the GLBA
  for guidance with regard to the reformulated question. However, the
  issue proves to be a novel one in the country since few courts,
  federal or state, have addressed the exceptions to the GLBA.''}

For our framework we need to take into account court hierarchies to
identify binding precedents. In common law jurisdictions, inferior
courts are bound by the decisions of superior courts; in civil law
jurisdictions superior courts usually have higher authority without
being strictly binding. In federal states like the USA or Germany we
need to account for parallel hierarchies on state and on federal
levels. This complex hierarchy has significant implications on stare
decisis.\cutShort{\footnote{For example, state courts in the United States are
  not considered inferior to federal courts but rather constitute a
  parallel court system. While state courts must follow decisions of
  the United States Supreme Court on questions of federal law, federal
  courts must follow decisions of the courts of each state on questions
  of that state's law. 
  }}

Hence, in our framework every case needs to be annotated by a court
which is part of a \emph{court hierarchy}, to identify the character
of precedents, binding or potentially persuasive.

\paragraph{Ratio decidendi and obiter dicta}
Regarding the court's decision text, we need to differentiate between
two types of statements. The actual binding property of a precedent
has only those statements and legal reasoning that are necessary for
the rationale of the decision. These necessary statements as called
{\it ratio decidendi} and constitute the binding precedent. Further
statements and reasoning that are not essentially necessary for the
decision are called {\it obiter dicta}. These are not binding but can
be referenced as persuasive precedents.

For our reasoning framework we need to differentiate and annotate
statements into these two different categories to correctly identify
binding precedents.



\section{Defining The \fworkName\ Framework}\label{sec:framework}

Reflecting the observations just made, we define cases
(Section~\ref{sec:cases}) and case law databases
(Section~\ref{sec:cld}). Thereby we also explain how to model the
legal principles described in Section~\ref{sec:ingredients}. Then, we
define how the database can be used in order to deduce facts outside
the framework (Section~\ref{sec:deducibility}). We analyze our
framework, validating a number of basic desirable properties of case
law databases (Section~\ref{sec:cldProp}). We finally show, for
privacy regulations specifically, that our framework matches the
requirements identified by previous work~\cite{barth2006privacy}
(Section~\ref{sec:privCases}).

Throughout this section, we assume an underlying logic in which world
properties are expressed and reasoned about. Our framework is
parametric with respect to the precise form of that logic. The
requirements the logic has to fulfill are interpreting predicates as
relations over objects, supporting universal truth/falseness (denoted
respectively as $\top$ and $\bot$), conjunction (denoted $\wedge$),
entailment (denoted $A\models B$ if formula $A$ entails formula $B$),
and monotonicity regarding entailment, i.e., if $A\models B$ then
$A\wedge C\models B$ for any formula $C$. \cutShort{We will discuss
  later on (Section~\ref{sec:automation}) a particular kind of logics
  suitable in our setting.} As an intuition when reading the
following, the reader may assume we are using a first-order predicate
logic.

\subsection{Introducing Cases}\label{sec:cases}
As we have seen, a case consists of a decision formula, a case
description, a court, and a proof tree. The first three components are
straightforward to capture formally (courts are represented by a
finite set $\courtSet$ of court identifiers). Designing the proof tree
is more involved since it needs to capture the judge's
justification. We distinguish between different kinds of nodes in the
tree depending on the role the respective statements play in the
justification: Does a sentence make an axiomatic statement, or form
part of the case description? Does it refer to a previous case,
adopting a decision under particular prerequisites? Does it make an
assessment on the truth of a particular statement (e.g., that a
particular piece of information is or is not to be considered private)
under particular prerequisites?\cutShort{ All such statements are
  ``standalone'' in the sense that they are not implications of
  previous arguments in the justification at hand.} We therefore
reflect \replShort{them}{these ``standalone'' statements} in the leaf
nodes of the proof tree, categorized by the three different types of
statements mentioned.

The inner nodes of the tree perform logical deductions from their
children nodes, representing the reasoning inherent in the
justification, i.e., the conclusions that are made until finally, in
the tree root, the decision formula is reached. \cutShort{Thereby,
  every inner node is annotated by an arbitrary formula.} We
differentiate between two kinds of reasoning steps, $\AND$-steps and
$\OR$-steps.The $\OR$-steps reflect the principle of \emph{independent
  decision grounds}\cutShort{, i.e., the cases that a judge increases
  legal certainty by listing arguments that all for themselves are
  sufficient for the conclusion}. The $\AND$-step is the natural
conclusion steps that is used to ensure that the decision made is
reached through the argumentation.

In order to avoid a recursive definition, we need a (possibly
infinite) set of case identifiers $\mathcal{C}_I$. Throughout the
paper we assume a fixed given set $\mathcal{C}_I$.\cutShort{ This
  leads to the following definition:}

\begin{definition}[Case]\label{def:case}
  A case $C$ is a tuple $(\confPred,\caseDesc,\prTree,\crt)$
  \replShort{such that}{s.t.}
 \vspace{-\topsep}\begin{itemize}
\item $\confPred$ is a formula that we call the \emph{decision formula}
  of $C$.
\item $\caseDesc$ is a formula describing the case's circumstances.
\item $\prTree$ is a (finite) tree consisting of formulas $f$ where
  the formula of the root node is $\confPred$.  Inner nodes are
  annotated with $\AND$ or $\OR$ and leaves are annotated with
  $l\in\{\labAxiom,\labDec\} \cup \{\labCase(i) \mid i\in\mathcal{C}_I
  \}$. Leaf formulas $l$ are additionally associated with a
  \emph{prerequisite} formula $\pre$. For leaves annotated with
  $\labAxiom$, we require that $\pre=l$.
\item $\crt\in\courtSet$.
\end{itemize}\vspace{-\topsep}

\noindent For leaf formulas $l$, we refer to $l$ as the node's
\emph{fact}, and we will often write these nodes as
$\pre\rightarrow\fact$ where $\fact=l$.

By the \emph{prerequisites} of an inner node $n$ with children nodes
$n_1,\ldots, n_k$, denoted as $\pres(n)$, we refer to $\bigvee_{1\leq
  i\leq k} \pres(n_i)$ if $n$ is annotated by $\OR$ and
$\bigwedge_{1\leq i\leq k} \pres(n_i)$ if $n$ is annotated by
$\AND$. The prerequisites of a case $C$ are the prerequisites of the
root node and denoted by $\pres_C$. We define analogously the
\emph{facts} of a node and a case. We will often identify formulas
with proof tree nodes. Given a case $C$, by $\confPred_C$ we denote
the decision formula of $C$.

Let $\bf{C}$ be a set of cases and $\mu:\bf{C}\to\mathcal{C}_I$ a
function. If for every reference $\labCase(i)$ in $\bf{C}$, there is
an $D\in\bf{C}$ with $\mu(D)=i$, we call the set $\bf{C}$
\emph{closed} under $\mu$.
\end{definition}

We assume \emph{world knowledge} common to all cases. In the example
of argumentation ends in Section~\ref{sec:ingredients}, it is assumed
that the reader knows that the predicate
$\textsf{is\_judical\_process}$ holds for any case. Formally, the
world knowledge is a formula $\KB_W$ (naturally, a conjunction of
world properties) in the underlying logic.

Definition~\ref{def:case} is purely syntactic, imposing no
restrictions on how the different elements are intended to behave. We
will fill in these restrictions one by one as part of spelling out the
details of our framework, forcing cases to actually decide a conflict
and behave according to the legal principles. One thing the reader
should keep in mind is that $\pre\rightarrow\fact$ is \emph{not}
intended as a logical implication. Rather, $\pre$ are the
prerequisites that a judge took into account when making the
assessment that $\fact$ (e.g., the privacy status of a piece of
information) is considered to be true under the circumstances
$\caseDesc\models\pre$.\cutShort{ The $\pre\rightarrow\fact$
  dependencies thus model the human element in case law, which we
  consider to be outside of what we can capture with formal logic.}
This solely captures human decisions such as trade-off decisions.
However, the frameworks allows reasoning about consequence of such
decisions.  The formulas $\pres_C$, and respectively $\facts_C$,
collect all prerequisites needed to apply the proof tree, and
respectively all facts needed to execute the proof tree; axiom leaves
act in both roles.

In principle, a case has the purpose to decide a formula
$\confPred$. However, while justifying that a formula holds, e.g.,
that a telecommunication company has to delete connection data after a
certain amount of time, the court might decide other essential
subquestions.\cutShort{ In the given example, this could be that
  connection data is personal data.} This concept is conveniently
captured through the notion of \emph{subcases}.

\begin{definition}[Subcase]
Let $C=(\confPred,\caseDesc,\prTree,\crt)$ be a case and $n\in\prTree$
a node. Let $\sub(n)$ be the subtree of $\prTree$ with root node $n$.
The case $\subcase(C,n) := (n,\caseDesc,\sub(n),\crt)$ is a
\emph{subcase} of $C$.
%
\end{definition}

Another aspect that is of interest when referencing cases is the
degree of abstraction. For example, one case could decide that a
specific telecommunication company $C$ has to delete connection
information $D$ of some user $U$ after a specific time period $t$.
The question of how this decision can be used in order to decide the
question for different companies $C'$ or different information $D'$ is
covered by the legal concept of material difference. For this work, we
assume that a judge specifies the allowed difference in the
prerequisites of a decision.\cutShort{ However, it could also be
  modeled by introducing metrics and thresholds when referencing
  (sub-)cases.}

Our definition of cases, so far, is generic in the sense that it may
be applied to any domain of law. To configure our framework to privacy
regulations more specifically, a natural approach is to simply
restrict the permissible forms of decision formulas. We explicitly
leave out legal domains such as individualized sentencing or measuring
of damages. Decisions in the privacy context are about whether or not
a particular action is legal when executed on particular data. We
capture this by assuming a dedicated predicate $\legalPred$, and
restricting the decision formula to be an atomic predicate of the form
$\legalPred(a)$, where $a$ is an action from an underlying set
$\actionSet$ of possible actions treated as objects (constants) in the
underlying logic. This can also be used in other legal domains, but it
turns out to be sufficient to connect our formalization of privacy
cases with other policy based approaches. Note that, in contrast to
other policy frameworks, we do not need to add the context to the
predicate, as the context is contained in the case, via nodes of the
form {\it ``if the transfer-action $a$ has purpose marketing and the
  receiver is a third party, then $\lnot\legalPred(a)$''}. As
decisions about the legality of actions are not naturally part of the
common world knowledge $\KB_W$, nor of the case description
$\caseDesc$ itself, our modeling decision is to disallow the use of
$\legalPred$ predicates in these formulas. In other words, the world
and case context describe the circumstances which are relevant to
determining action legality, but they do not \cutShort{themselves}
define whether or not an action is legal. \cutShort{This yields the
  following definition:}

\begin{definition}[Privacy Case]
  Given world knowledge $\KB_W$ and action set $\actionSet$, a case
  $C=(\confPred,\caseDesc,\prTree,\crt)$ is a \emph{privacy case} if
  $\confPred\in\{\lnot \legalPred(a),\legalPred(a)\}$ for some action
  $a\in\actionSet$, where the $\legalPred$ predicate is not used in
  either of $\KB_W$ or $\caseDesc$.
\end{definition}

Starting to fill in the intended semantics of cases\cutShort{, i.e.,
  of the structures allowed as per Definition~\ref{def:case}}, we
first capture the essential properties a case needs to have to ``make
sense'' as a stand-alone structure. Additional properties regarding
cross-case structures will be considered in the next subsection. We
will use the word ``consistency'' to denote this kind of property. The
following definition captures the intentions behind cases:

\begin{definition}[Case Consistency]\label{def:case:consist}
  Let $C=(\confPred,$ $\caseDesc,$ $\prTree,\crt)$ be a case. $C$ is
  \emph{consistent} if the following holds (for all nodes $n$ where
  $n_1,\ldots,n_k$ are its child nodes)

\noindent\begin{tabular}{p{.5\textwidth}l}
  (i) $\KB_W\wedge\caseDesc\not\models\bot$ & (ii) $\KB_W\wedge\caseDesc\models\pres_C$\\
  (iii) $\KB_W\wedge\caseDesc\wedge\facts_C\not\models\bot$ &\\
  (iv) $\bigwedge\limits_{1\leq i\leq k}n_i\models n$ 
  if $n$ is an $\AND$ step &
  and $\bigvee_{1\leq i\leq k}n_i\models n$ if $n$ is an $\OR$ step
\end{tabular}
\end{definition}

Regarding (i), if the world knowledge contradicts the case
description, i.e., $\KB_W\wedge\caseDesc\models\bot$, then the case
could not have happened\cutShort{ in reality}. Similarly, (iii) the
case context must not contradict the facts that the proof tree makes
use of (this subsumes (i), which we kept as it \replShort{makes the
  definition more readable}{improves readability}). As for (ii), the
case context must imply the axioms as well as the prerequisites which
the present judge (assessments) or other judges (references to other
cases; see also Definition~\ref{def:cld:ref}) assumed to conclude
these facts. (iv) says that inner nodes must represent conclusions
drawn from their children\cutShort{ (remember here that $n_i$, for
  leaf nodes $\pre\rightarrow\fact$, refers to $\fact$)}.

The $\OR$ nodes of the proof tree reflect the legal argumentation
structure of \emph{independent decision grounds}, the judge gives
several arguments\cutShort{, each of which is sufficient}. If the
judge of a later case decides that one of these arguments is invalid
for the conclusion, he needs to be able to falsify only one of the
branches and not the whole tree. \cutShort{In other words, the tree structure
gives ``syntactic sugar'' that makes it possible to reflect the
justification more closely and thereby marks which subsets of leaf
nodes are sufficient in order to reach decision $\confPred$.}



\subsection{Combining Cases to Case Law Databases}\label{sec:cld}
The quintessential property of case law is that cases make references
to other cases. These references are necessary to formulate several
legal principles\cutShort{ of Section~\ref{sec:ingredients}}.

The legal principles \emph{false material difference} and
\emph{reversing decisions} define requirements for when not to
reference a case, either because it contains a mistake or because the
opinion has changed over time. Therefore, we consider the design
cleaner if both principles are covered by the same mechanism of the
framework\replShort{. There are several options to model the
  principles: first, the reversed decision could be covered by time,
  i.e., by a requirement to refer to the newest case that is
  applicable regarding the circumstances. However, the false material
  difference cannot be covered by that. Another approach is to}{ and
  hence we} denote single $\labDec$ nodes as unwarranted, i.e., to
forbid the reference to be used thereafter. \cutShort{This solution
  can model both principles \emph{false material difference} and
  \emph{reversing decisions}. We explicitly decided to model the
  mechanism of unwarranted nodes outside of the cases. Assume a case
  would decide that another decision was unwarranted. This leads to
  another decision that could potentially be marked as unwarranted
  later on implying that it is again correct to cite the
  case. Consequently, this would lead to a set of time intervals
  during which the citation of nodes is warranted. However, after
  legal consultation we figure out that this complication does not
  meet practice, i.e., once a decision is unwarranted it will not
  become warranted again; hence we simplified the mechanism.}

We require a different mechanism to differentiate cases we must agree
with and cases which we may use as reference. Unwarranting rather
defines which decisions must not be referenced. In particular, we need
to differentiate between assessments coming from the legal principles
\emph{ratio decidendi} and \emph{obiter dicta}. While the part of the
decision following \emph{ratio decidendi} leads to a binding
precedent, the \emph{obiter dicta} part is not binding.  Thus, we
introduce predicates $\mayref$ and $\mustref$. It also provides a
mechanisms to respect the \emph{court hierarchy}. Intuitively,
$\mayref(C_1,C_2)$ denotes the circumstances that case $C_1$ may
reference case $C_2$; $\mustref(C_1,C_2)$ analogously denotes that
$C_1$ must agree with $C_2$.

In addition, we need to introduce the concept of time by a total order
$\leq_t$ over cases. This concept allows us to formulate the
requirement that references can only point to the past.\cutShort{ Using all
these constructs, we can define a case law database.}

\begin{definition}[Case Law Database (CLD)]
  A \emph{case law database} is a tuple
  $\DB=({\bf{C}},\leq_t,\mustref,\mayref, \mu, U)$ such that:
\vspace{-\topsep}\begin{itemize}
\item $\bf{C}$ is a set of cases. We will also write $C\in\DB$ for $C\in\bf{C}$.
\item $\mu:\bf{C}\to\mathcal{C}_I$ is an injective function such that
  $\bf{C}$ is closed under $\mu$. In the following we will also write
  $\labCase(D)$ for $\labCase(i)$ if $\mu(D)=i$.
\item Let $\leref\ := \{(C,D)\mid D$ contains a $\labCase(C)$ node$\}$
  and $\leq_t$ is an order that we call \emph{time order} of the
  cases.  It has to hold:
\vspace{-\topsep}
\begin{center}\begin{tabularx}{.5\textwidth}{rl}
$\mustref\subseteq$ & \multirow{2}{*}{$\mayref\subseteq\leq_t\subseteq\bf{C}\times\bf{C}$}\\
$\leref\subseteq$
\end{tabularx}
\end{center}
\item $U$ specifies the unwarranted nodes, i.e., $U:\bf{C}\to\bf{N}$
  is function such that
  \begin{itemize}
  \item $\bf{N}$ is a subset of the nodes labelled with $\labDec$ or
    $\labCase$ in the cases $\bf{C}$.
  \item The set increases monotonic, i.e., $C\leq_t D\implies U(C)\subseteq U(D)$.
  \end{itemize}
  We denote the \emph{unwarranted} nodes of $\DB$ by $U(\DB) :=
  \bigcup_{C\in\bf{C}}U(C)$.
\end{itemize}
%
%
\end{definition}

The function $\mu$ is used to remove the recursive definition of a
case and enables us to connect cases via their individual semantics.

Regarding the relations $\mustref$ and the $\mayref$ we made two
design decisions. First, we require to not link $\mustref$ and the
actual references $\leref$. On the one hand, there might be precedents
which are not applicable, but on the other hand, we want the freedom
to define $\mustref$ and $\mayref$ only depending on the court
hierarchy\cutShort{, i.e., independent of the satisfaction of some
  precedent's preconditions}. The second design decision is to base
these relations on cases instead of decision nodes. As for the first
decision, the purpose is to make an instantiation of the definition
only depending on the court, but we need to be careful regarding the
principles \emph{ratio decidendi} and \emph{obiter dicta}. Since one
of them is not binding, i.e., a $\mustref$ and the other is. This
differentiation can be achieved by replacing every case with a set of
cases. We require this to be part of the modeling
process.\cutShort{However, it is possible to automatically identify
  parts of the proof that are optional to reach the final decision in
  the root node.}

We did not add further restrictions since they may depend on local
law. For example, there is a \emph{vertical stare decisis} in US law,
implying that higher court decisions have to be considered. There is
also the term of \emph{horizontal stare decisis} that requires
respecting siblings in the hierarchy. This principle does not
necessarily hold, but is under discussion. However, the definition of
must- and may-references allows modeling both.

\begin{example}[Must-agree and may-references for a court
  hierarchy]\label{example:mustmay}
  Assume the set of courts $\courtSet$ is partially ordered by
  $\leq_{\S}$, i.e., there is a court hierarchy. In this case, we
  could model $\mustref$ by
\begin{tabular}{rl}
  $\mustref$ = &$\{ (C_1,C_2) \mid C_i=(\confPred_i,d_i,p_i,\crt_i),i\in\{1,2\}, C_1\leq_t C_2,$\\ 
  &and $\crt_1\leq_{\S}\crt_2 \}$.
\end{tabular}

It is easy to see that the $\mustref$ predicate actually only depends
on the $\crt$ and not on the other parameters of the proof. We call
this property \emph{court-dependency}. 
\end{example}

The key property of unwarranted decisions is that they are time
dependent. In order to only use warranted decisions when referencing,
we define warranted subcases as follows:

\begin{definition}[Warranted Subcase]\label{def:war:sub}
  A subcase $(\confPred, \caseDesc,\prTree,\crt)$ is \emph{warranted}
  with respect to a set $N$ of nodes if the case $(\confPred,
  \caseDesc, \prTree', \crt)$ is consistent where $\prTree'$ is
  derived from $\prTree$ by replacing every precondition of a node
  $n\in N$ by
  $\bot$. 
\end{definition}

It remains to define when a case law database can be considered to be
consistent. To that end, we consider case references and conflicts
between cases. Starting with the former, we obtain:

\begin{definition}[Correct Case Reference]\label{def:cld:ref}
  Let $\DB$ be a case law database and $C =
  (\confPred,\caseDesc,\prTree,\crt)$ a case in $\DB$. A leaf node
  $\pre\rightarrow\fact$ in $\prTree$ annoted with $\labCase(D)$
  \emph{references correctly} if $D_u = (\fact,$ $\caseDesc_D,$
  $\prTree_D,$ $\crt_D)$ is a warranted subcase of a case $D\in\DB$
  w.r.t. $U(C)$, $\mayref(C,D)$ holds and
  $\KB_W\wedge\pre\models\pres_D$. $C$ \emph{references correctly} if
  all its leaves annoted with $\labCase(D)$ reference correctly.
\end{definition}

Consider that, when referencing a (sub)case $D$ as
$\pre\rightarrow\fact$ from our case $C$ at hand, we are essentially
saying that the same argumentation applied in $D$ can be applied in
our case, to prove $\fact$ under circumstances $\pre$. So we need to
show that this applicability of arguments is actually given. This is
ensured by $\KB_W\wedge\pre\models\pres_D$ because $\pres_D$ collects
all prerequisites, axioms and otherwise, needed to apply $D$. Note
that, if $C$ is consistent, by Definition~\ref{def:case:consist} (ii)
it holds that $\KB_W\wedge\caseDesc\models\pre$ and thus
$\KB_W\wedge\caseDesc\models\pres_D$. \cutShort{Note further that
  $\KB_W\wedge\pre\models\pres_D$ defines the role of $\pre$ as
  providing a condition sufficient to entail ``the other judge's
  prerequisites''.} As the same applies recursively to the case
references made in $D$, we know that $\pre$ (given $\KB_W$ and
$\caseDesc$) 
entails \emph{all} judge decisions underlying the assessment
$\fact$. \cutShort{We will formalize this in
Theorem~\ref{theorem:allFollowsFromDecisions}.
}

We are now almost in the position to define consistency \cutShort{at
  the 
  level} of the entire case law database. The last missing piece in
the puzzle is to identify when cases should be considered to be in
conflict --- which naturally occurs in case law databases where
\cutShort{different} judges may make different decisions. We capture
this through pairs of cases whose prerequisites are compatible, while
their facts are contradictory:

\begin{definition}[Case Conflict]\label{def:conflict}
  Let $C_1$ be a case in $\DB$ and $C_2$ be a warranted case
  w.r.t. $U(C_1)$. We say that $C_1$ is \emph{in conflict with} $C_2$
  if and only if

\begin{tabular}{p{.5\textwidth}l}
  (i) $\KB_W\wedge\pres_{C_1}\wedge\pres_{C_2}\not\models\bot$ &
  (ii) $\KB_W\wedge\facts_{C_1}\wedge\facts_{C_2}\models\bot$\\
  (iii) $\mustref(C_1,C_2)$
\end{tabular}

\noindent A case $C$ is in conflict with $\DB$ if there is a $D\in\DB$
s.t. $C$ is in conflict with $D$.
\end{definition}

We ignore the case descriptions here, other than what is explicitly
employed as axioms in the proof trees: we consider cases to be in
conflict if one \emph{could} construct a case (e.g.,
$\pres_{C_1}\wedge\pres_{C_2}$) which would make it possible to come
to a contradictory decision.
%
%
We define case law database consistency as follows:

\begin{definition}[Case law database consistency]
  A case law database $\DB=({\bf{C}},\leq_t,\mustref,\mayref, \mu, U)$ is
  \vspace{-\topsep}\begin{itemize}
  \item[(i)] \emph{case-wise consistent} if every $C\in\DB$ is consistent,
  \item[(ii)] \emph{referentially consistent} if every $C\in\DB$
    references correctly, and
  \item[(iii)] \emph{hierarchically consistent} if every $C\in\DB$ is
    not in conflict with $\DB$.
  \item[(iv)] \emph{warrants consistently} if for every $C$ holds:
    $U(C)$ contains all $\labCase(D)$ nodes where $D$ is an
    unwarranted subcase w.r.t. $U(C)$.
  \end{itemize}\vspace{-\topsep}
  We call $\DB$ \emph{consistent} if it warrants consistently and is
  hierarchically, referentially and case-wise consistent.
\end{definition}



\subsection{Deriving Legal Consequences: Deducibility and
  Permissibility}\label{sec:deducibility}
In the following we assume that the predicates $\mayref$ and
$\mustref$ of the $\DB$ do not depend on the case description, the
decision formula or the proof tree, but are only court dependent,
cf. Example~\ref{example:mustmay}. As a consequence, we know the value
of these predicates for formula values and case descriptions which are
not contained as a case in the database given only the court level of
the case. In other words, we require an operation $\DB\cup\{C\}$ that
puts $C$ at the end of the timeline regarding $\leq_t$, assigns a
fresh identifier $i\in\mathcal{C}_I$ to $C$ with $\mu$, uses as
$U(C):=U(\DB)$, and adopts $\mustref,\mayref$ appropriately and is
independent of the decision formula and the proof tree. This operation
is needed to apply the framework to situations not contained in the
database.

Obvious applications of our framework are advanced support for case
search\cutShort{ (based on logic operations over the case
  descriptions, decision formulas, etc.)}, and consistency
checking\cutShort{ (given a case $C$, is $C$ consistent and does it
  reference correctly?)}. A more advanced task is to evaluate the
legality of actions given the cases reflected in the database. For
example, when designing a course administration system, one may ask
``Am I allowed to store students' grades in the system?''  Our
formalism supports this kind of question at different levels of
strength, namely:

\begin{definition}[Deducibility and Permissibility]\label{def:deduc}
  Let $\DB=({\bf{C}},\leq_t,\mustref,\mayref, \mu, U)$ be a consistent
  CLD, and $f$ a formula. We say that $f$ is \emph{permitted} in $\DB$
  under circumstances $\caseDesc$ and court $\crt$ if there exists a
  case $C=(f,\caseDesc,\prTree,\crt)$ such that $\prTree$ does not
  contain nodes labeled with $\labDec$, and $\DB\cup\{C\}$ is
  consistent (where $C$ is inserted at the end of the timeline
  $\leq_t$). We say that $f$ is \emph{uncontradicted} in $\DB$ under
  $\caseDesc$ and $\crt$ if $\lnot f$ is not permitted under
  $\caseDesc$ and $\crt$. We say that $f$ is \emph{deducible} if it is
  permitted and uncontradicted.

For sets $F$ of formulas, we say that $F$ is permitted in $\DB$ under
$\caseDesc$ and $\crt$ if there exists a set of cases
$\{C_f=(f,\caseDesc,\prTree_f,\crt) \mid f \in F\}$ such that every
$\prTree_f$ does not contain nodes labeled with $\labDec$, and
$\DB\cup\{C_f \mid f \in F\}$ is consistent (where the $C_f$ are
inserted in any order at the end of the timeline $\leq_t$).
\end{definition}

It might be confusing at first why we attach to $f$ the weak attribute
of being ``permitted'' if we can construct a case supporting it. The
issue is, both $f$ and $\neg f$ may have such support in the same
database. This follows directly from the freedom of different courts
to contradict each other. If two courts at the same level decide
differently on the same issue, then that is fine by our
assumptions
.  Hence, to qualify a formula $f$ for the strong attribute of being
``deducible'', we require the database to permit $f$ and to not permit
its contradiction.

\cutShort{Note that permissibility and deducibility are 
also dependent on the circumstances $\caseDesc$ and the court
$\crt$. For example, when we answer ``was it legal to send data $D$ to
party $P$?'', it matters for which purpose the data was sent. That
information is contained in the $\caseDesc$. The court level has
several interpretations here: the court might be chosen to match the
local court of the party asking the question. But the court level can
also be viewed as a level of confidence. Permissibility is a
``stronger'' guarantee for lower court instances, because we can then
deduce without incurring conflicts to instances higher up. Hence lower
court instances can be used to obtain permissibility ``with high
confidence'', and contradictions ``with low confidence''
. Vice versa, higher court instances can be used to obtain
permissibility ``with low confidence'' and contradictions ``with high
confidence''. 
}

The concept of deducibility of a \emph{set} $F$ of formulas is
interesting because, in general, this is not the same as deducing each
formula in separation. In particular, while each of $f$ and $\neg f$
may be permitted in the same database, $\{f, \neg f\}$ is never
permitted because adding the hypothetical supporting cases necessarily
incurs a hierarchical conflict. Permissibility of $F$ is also
not the same as permissibility of $\bigwedge_{f \in F} f$ because the
latter makes a stronger assumption: all cases referred to in order to
conclude $\bigwedge_{f \in F} f$ must have compatible
prerequisites. So deducibility of formula sets forms a middle ground
between individual and conjunctive deducibility.

\begin{theorem}\label{thm:deducSets}
  There is a consistent case law database $\DB$, case description
  $\caseDesc$ and court $\crt$, such that there is a set $F$ of
  formulas for each of the following properties (in $\DB$ under
  circumstances $\caseDesc$ and court $\crt$):
  \vspace{-\topsep}\begin{enumerate}[(i)]
  \item For every $f\in F$, $f$ is permissible and $F$ is not
    permissible.
  \item $F$ is permissible, but $\bigwedge_{f\in F}f$ is not
    permissible.
  \end{enumerate}\vspace{-\topsep}
\end{theorem}
This theorem's proof and the details of all other proofs are given in
the Appendix\cutShort{~\ref{sec:proofs}}.

\paragraph{Characterizing Deducibility}
Deducibility is the central concept for answering questions that are
not explicitly answered by the database. However,
Definition~\ref{def:deduc} does not give an algorithmic description of
how to decide whether some formula is deducible. It is also
inconvenient for proving properties about permissibility and
deducibility. Thus, we give an equivalent characterization in the
following.

Intuitively, a formula should be permissible if there is a set of
warranted decisions which allow us to conclude the predicate and a
formula $f$ should be deducible if in addition no set of decisions
contradicts $f$. We will first define \emph{supporting sets} and then
prove that the intuition matches the definitions of permissibility and
deducibility.

\begin{definition}[Supporting set]\label{def:sup:set}
  Let $\DB=({\bf{C}},\leq_t,\mustref,\mayref, \mu, U)$ be a consistent
  case law database, $f$ a formula, $\caseDesc$ a case description and
  $\crt$ a court.  A set $\mathcal{A}$ of leaf nodes in $\DB$ that are
  labeled with $\labDec$ is a \emph{supporting set} for formula $f$ if
  the following holds: \vspace{-\topsep}\begin{enumerate}[(1)]
  \item\label{def:sup:set:p1}
    $\KB_W\wedge\caseDesc\models\bigwedge_{(\pre\rightarrow\fact)\in\mathcal{A}}
    \pre$
  \item\label{def:sup:set:p2}
    $\KB_W\wedge\caseDesc\wedge\bigwedge_{(\pre\rightarrow\fact)\in\mathcal{A}}
    \fact\models f$
  \item\label{def:sup:set:p3}
    $\KB_W\wedge\caseDesc\wedge\bigwedge_{(\pre\rightarrow\fact)\in\mathcal{A}}
    \fact\not\models\bot$
  \end{enumerate}

  \noindent A supporting set is unwarranted if it contains an
  unwarranted node w.r.t. any $C\in\bf{C}$. If it is not unwarranted
  it is warranted.

  \noindent A supporting set is \emph{consistent} with $\DB$ if $\DB\cup\{(\top,
  \caseDesc, \prTree, \crt)\}$ is consistent, where $\prTree$ consists
  of a root node with annotation $\top$ and leaf nodes with annotation
  $\labCase(C_n)$ for $n\in\mathcal{A}$, where $C_n$ is the case that
  contains node $n$.
\end{definition}

Note that a supporting set that is consistent with the $\DB$ leads to
consistency, and correct referencing, and does not create any conflicts.
The properties required in the definition are a consequence of the
definition of database consistency. A case constructed from a
supporting set would simply refer to all decisions and place the
formula at the root. \cutShort{Case consistency requires the properties
(\ref{def:sup:set:p1})-(\ref{def:sup:set:p3}) to hold; referential
consistency requires that the referenced leaf nodes are warranted and
hierarchical consistency requires that the supporting set is not in
conflict with $\DB$.}

The following theorem characterizes permissibility and deducibility
using supporting sets. This characterization suggests an algorithmic
way of deciding the properties and gives a tool for proving properties
about case law databases.

\begin{theorem}\label{theorem:allFollowsFromDecisions}
  Let $\DB$ be a consistent case law database, $f$ a formula,
  $\caseDesc$ a case description and $\crt$ a court. The following
  holds:
  \vspace{-\topsep}\begin{enumerate}
  \item\label{th:suppset:p1} $C\in\DB$ with warranted node $f$
    $\Rightarrow \exists \mathcal{A}$ that supports $f$
  \item $f$ is permitted (under circumstance $\caseDesc$ and court
    $crt$) $\Leftrightarrow \exists \mathcal{A}$ that supports $f$, is warranted, and
    is consistent with $\DB$
  \item $f$ is deducible $\Leftrightarrow \exists \mathcal{A}$ that
    supports $f$ and is consistent with $\DB$, and $\forall
    \mathcal{B}$ it holds that $\mathcal{B}$ does not support $\lnot f$,
    is unwarranted, or is not consistent with $\DB$
  \end{enumerate}

\end{theorem}


\subsection{General Properties of Case Law Databases}\label{sec:cldProp}
Introducing a new framework always comes with the risk of modeling
errors. A method for alleviating that risk is to prove properties that
the framework is expected to have. 
In order to validate the framework introduced here, we have proven
that (i) case references do not influence decisions
(Theorem~\ref{theorem:allFollowsFromDecisions}); in this subsection we
additionally prove that (ii) consistency is necessary for property (i)
(Theorem~\ref{theorem:consistencyNecessary}), and that (iii) neither
$\bot$ nor $\{f, \neg f\}$ are ever permitted
(Theorem~\ref{theorem:botBringsConflicts}).

Regarding (i), we have shown that every formula $f$ in the database
can be derived from a supporting set of previous decisions
(Theorem~\ref{theorem:allFollowsFromDecisions}) with the case
description and world knowledge. Hence there is no possible interplay
between case references that would make it possible to prove something
not 
backed up by judges' decisions.


Regarding (ii), Theorem~\ref{theorem:allFollowsFromDecisions} implies
immediately that, whenever a formula $f$ is deducible, then it follows
from decisions made by judges in previous cases. It is easy to verify
that our restrictions are necessary to ensure this, i.e., that this
property gets lost if we forsake either case-wise or referential
consistency:

\begin{theorem}\label{theorem:consistencyNecessary}
Let $\DB$ be a case law database, and let $f$ be any formula that does
not entail $\bot$. Then there exist cases $C_1$ and $C_2$, each with
root node $f$ and the empty case desc $\top$, such that (inserting
$C_i$ at the end of the timeline $\leq_t$):
\vspace{-\topsep}\begin{itemize}
\item If $\DB$ is case-wise consistent, then so is $\DB\cup\{C_1\}$.
\item If $\DB$ is referentially consistent, then so is
  $\DB\cup\{C_2\}$.
\item If there is a $\crt$ such that $\mustref(\crt)=\emptyset$, then
  in addition this holds: for each of $i = 1,2$, if $\DB$ is
  hierarchically consistent, then so is $\DB\cup\{C_i\}$.
\end{itemize}
\end{theorem}

We remark that, by restricting the formula $f$ only slightly, the
proof of Theorem~\ref{theorem:consistencyNecessary} can be
strengthened so as not to have to rely on a maximal court for ensuring
hierarchical consistency. In particular, if $f$ is made of predicates
that do not occur anywhere in the case law database, then the cases
$C_1$ and $C_2$ as constructed cannot be in conflict with any other
cases, thus preserving hierarchical consistency for arbitrary courts
$\crt$.
We finally prove (iii), non-permissibility of either $\bot$ or $\{f,
\neg f\}$:

\begin{theorem}\label{theorem:botBringsConflicts}
  The formula $\bot$ is not permitted in any case law database $\DB$,
  under any circumstances $\caseDesc$ and court $\crt$.  The same
  holds for $\{f, \neg f\}$ if $\crt\in\mustref(\crt)$.
\end{theorem}



\subsection{Privacy Cases and Norms}\label{sec:privCases}
We now point out an interesting property of privacy cases, and of case
law databases consisting only of privacy cases. We call such databases
\emph{privacy case law databases}.

Rule based privacy policies are a well established and widely used
concept. The rules that are used are usually reflected by norms
defining privacy regulations. However, neither rules nor norms are
reflected in the case law framework. In this subsection, we show that
we can use a natural definition of norms that can be extracted from
privacy cases. In addition, it is possible to transform a privacy case
to a normal form such that a norm that decides the case is
represented. \cutShort{Consequently, we also consider norm extraction
  as a reasoning task in Section~\ref{sec:reason:task}.}

At the core of privacy regulations are positive and negative norms, as
introduced by~\cite{barth2006privacy}. Positive norms are permissive
in the sense that they describe conditions that allow transactions
with personal data ($\phi\Rightarrow\legalPred(a)$). Negative norms,
in contrast, define necessary conditions for such transactions, i.e.,
they forbid transactions with personal data unless certain conditions
are met ($\phi\Rightarrow\lnot\legalPred(a)$). \cutShort{We formulate
  negative norms as conditions that lead to the denial of
  transactions.}

\begin{definition}[Norms]\label{def:norm}
  Let $a\in\actionSet$.  A \emph{norm} is a formula that has the form
  $\phi\Rightarrow p$ where $\legalPred(a)$ does not occur in
  $\phi$. The norm is a positive norm, denoted $\phi^+$, if
  $p=\legalPred(a)$ and a negative norm, denoted $\phi^-$, if
  $p=\lnot\legalPred(a)$. 
%
  A norm $\phi$ decides $p$ given $f$ if $\KB_W\wedge f\models \phi$.
\end{definition}

In the case law framework, norms are hidden by judges' assessments.
However, in the spirit of
Theorem~\ref{theorem:allFollowsFromDecisions}, norms are reflected by
sets of cases that could be referenced in order to support either the
legality of an action (positive norm) or its illegality (negative
norm).  In the following theorem, we show that we can extract a norm
for every privacy case avoiding the recursion of
Theorem~\ref{theorem:allFollowsFromDecisions}.

\begin{theorem}\label{theorem:norm:extract}
  Let $\DB$ be a consistent privacy case law database and
  $C=(\confPred,\caseDesc,\prTree,\crt)\in\DB$. Then there is a norm
  $\phi$ that decides $\confPred$ given $\caseDesc$. In particular,
  there are formulas $\phi_W,\phi_S$ such that $\legalPred(a)$ does not
  occur in these formulas and
\begin{tabular}{p{.5\textwidth}l}
(1) $\facts_C\Rightarrow \phi_W\wedge(\phi_S\Rightarrow\confPred)$ & 
(2) $\phi_W\wedge(\phi_S\Rightarrow\confPred)\Rightarrow\confPred$ \\
\end{tabular}

\end{theorem}

The formulas $\phi_W$ and $\phi_S$ can be used to construct a normal
form of privacy cases. In particular, this normal form is consistent and
allows reading off norms. 

\begin{corollary}[Normal forms]\label{cor:norm:forms}
  Let $\DB=({\bf{C}},\leq_t,\mustref,\mayref, \mu, U)$ be a privacy case law database,
  $C=(\confPred,\caseDesc,\prTree,\crt)\in\DB$ be a case, and $D$ be
  the set of $C$'s leaf nodes. $N(C)$ is the case that consists of a
  root node $\confPred$, two inner nodes $\phi_w$ and
  $\phi_S\Rightarrow\confPred$ and the leaf nodes $D$ as children of
  both inner nodes. We call $N(C)$ the \emph{normal form} of $C$.
  If $\DB$ is consistent, then
  $({\bf C}\backslash\{C\}\cup\{N(C)\},\leq_t)$ is also consistent (where
  $N(C)$ is placed at the position of $C$ w.r.t. $\leq_t$).
\end{corollary}

In order to define $N(C)$, we need to duplicate the leaf nodes since
the transformations to get $\phi_W$ and $\phi_S$ ignore which $\fact$
is needed to get the corresponding formula. Thus, a leaf node's
$\fact$ could end up in both formulas $\phi_W$ and $\phi_S$.

In conformance with~\cite{barth2006privacy}, we can conclude from %
deducibility of an action that there is a positive norm supporting it
and show that no negative norm can be applied, i.e., all negative
norms are respected (Theorem~\ref{theorem:botBringsConflicts}).  



\section{Reasoning Tasks}\label{sec:reason:task}

We now discuss the reasoning tasks associated with our framework ---
how to answer questions such as ``are we allowed to send data $D$ to
some party $P$?'' --- in more detail, giving an algorithm sketch and
brief complexity analysis (in terms of the number of reasoning
operations required) for each.

\paragraph{Consistency} Analyzing and keeping the state of the case
law database consistent is of vital importance for its usefulness;
cf.\ Theorem~\ref{theorem:botBringsConflicts}. As in the definition of
consistency, we split the task of checking consistency into case-wise,
referential, and hierarchical consistency. Due to their simplicity, we
postpone the detailed description of their algorithms to the appendix.

All of these properties are defined per case, i.e., the case wise
check of the corresponding property has to be repeated $|\DB|$
times. Following the respective definition, checking case consistency
costs 
$|\prTree+1|$ entailment operations 
and checking correct referencing for $C$ costs $\refer(C)$ where
$\refer(C)$ is the number of nodes in $C$ annotated by
$\labCase(D)$. Hierarchical consistency can be checked along the time
line $\leq_t$ only testing for conflicts with earlier cases. So for
the $i$-th case, we need at most $(i-1)\cdot 2$ entailment checks,
since every conflict check requires $2$. Consequently, we require
$|\DB|\cdot(|\DB|+1)$ entailment checks.

\cutShort{The property whether the case law database warrants
  consistently can be checked using one entailment test per reference
  to a subcase containing an unwarranted decisions node.}

\paragraph{Deducibility and Permissibility} As deducibility amounts to
two consecutive permissibility checks, we consider the latter
exclusively. We are given a database $\DB$, a formula whose
permissibility should be checked, as well as a case description
$\caseDesc$ and a court $\crt$ forming the circumstances.
\replShort{By Theorem~\ref{theorem:allFollowsFromDecisions},
  p}{P}ermissibility is equivalent to the existence of a supporting
set $\mathcal{A}$ for $f$ that is consistent with \replShort{the
  database}{$\DB$}. Thus the task of permissibility\cutShort{, i.e.,
  giving a ``yes'' vs.\ ``no'' answer,} can be reduced to checking the
existence of a suitable set $\mathcal{A}$. If the answer is ``yes'',
we can also output a witness, i.e., a hypothetical case $C$ showing
permissibility. A straightforward means for doing this is to set $C :=
(f,\caseDesc,\prTree,\crt)$ where $\prTree$ consists of root node $f$,
one leaf node $l$ labeled with $\labCase(D)$ for every $D \in
\mathcal{A}$, as well as one leaf node $\KB_W\wedge\caseDesc$ labeled
with $\labAxiom$. For convenience, we will denote this construction by
$C(\mathcal{A})$. 
See Algorithm~\ref{alg:permissibility}.

\renewcommand\topfraction{0.85}
\renewcommand\bottomfraction{0.85}
\renewcommand\textfraction{0.1}
\renewcommand\floatpagefraction{0.85}
\setlength\floatsep{.5\topsep}
\setlength\textfloatsep{.5\topsep}
\setlength\intextsep{.5\topsep}
\IncMargin{1em}
\RestyleAlgo{boxruled}
\LinesNumbered
\begin{algorithm}[htb]
  \Indm
  \SetKwInOut{Input}{Input}
  \SetKwInOut{Output}{Output}
  \caption{Permissibility\label{alg:permissibility}}
  \Input{A formula $f$, case description $\caseDesc$, court $\crt$, and
    a consistent CLD $\DB$}%
  \Output{A case $C=(f,\caseDesc,\prTree,\crt)$ such that $\DB\cup\{C\}$
    is consistent (where $C$ is set to be the maximum w.r.t. $\leq_t$),
    or $\bot$ if no such $C$ exists} %
  \Indp Test whether $\KB_W\wedge\caseDesc\models\bot$. If so, output $\bot$.\\
  Test whether $\KB_W\wedge\caseDesc\models f$. If so, output
  $(f,\caseDesc,\prTree,\crt)$ where $\prTree$ is the proof tree
  consisting of a leaf node labeled by $\labAxiom$ containing $f$.\\
  Set $\mathcal{N}:=\emptyset$.\\
  \For{every $D\in\DB$ and every $(\pre\rightarrow\fact)\in D$ labeled
    $\labDec$}{%
    Check if $\KB_W\wedge\caseDesc\models\pre$\\
    Check if $\KB_W\wedge\caseDesc\wedge\concl\not\models\bot$\\
    If both checks succeed, set $\mathcal{N} :=
    \mathcal{N}\cup\{(\pre\rightarrow\fact)\}$.
  }\label{enum:correct:ref}%
  \For{$\mathcal{A}\in 2^{\mathcal{N}}$\label{algo:ded:test}}{ Check
    that
    $\KB_W\wedge\caseDesc\models\bigwedge_{(\pre\rightarrow\fact)\in\mathcal{A}}\pre$\\
    Check that
    $\KB_W\wedge\caseDesc\wedge\bigwedge_{(\pre\rightarrow\fact)\in\mathcal{A}}\fact\models
    f$\\
    Check that
    $\KB_W\wedge\caseDesc\wedge\bigwedge_{(\pre\rightarrow\fact)\in\mathcal{A}}\fact\not\models\bot$\\
    \For{every $E\in\DB$ with $\crt\lecourt\crt_{E}$}{%
      Check that $E$ and $C(\mathcal{A})$ are not in conflict
      (cf. Algorithm~\ref{alg:hierarchical:consistent}).%
    }%
    If all three tests succeed, go on with step~\ref{algo:ded:end},
    otherwise continue with the next $\mathcal{D}$.  } %
  If a set $\mathcal{A}$ succeeded, output $C(\mathcal{A})$, otherwise
  output $\bot$.
  \label{algo:ded:end}

\end{algorithm}
 \DecMargin{1em}

 The correctness of the algorithm is shown by
 Theorem~\ref{theorem:allFollowsFromDecisions}\cutShort{; lines 10-12
   check that the set supports $f$ and lines 13-15 ensure that it is
   consistent with the database}.
 In contrast to our previous algorithms, deducibility checking as per
 Algorithm~\ref{alg:permissibility} requires an exponential number of
 entailment checks in the worst case\cutShort{ (a trivial bound is in
   the order of $2^{N}$ where $N$ is the number of decision nodes in
   the database)}. This raises the questions (1) whether or not this
 exponential overhead is inherent in the complexity of deciding
 permissibility, and (2) whether it is possible to encode the
 permissibility test directly into the logic
 instead. 
\cutShort{ In what follows, we shed some light on (1) and (2).}

The answer to (1) is a qualified ``yes'' in the sense that
permissibility checking essentially pre-fixes entailment checks with
an existential quantifier. As entailment checks correspond to
universal quantification, this intuitively means that for
permissibility we need to test the validity of a $\exists\forall$
formula, instead of a $\forall$ formula for entailment. So we add a
quantifier alternation step, which typically does come at the price of
increased complexity. This line of thought also immediately provides
an intuitive answer to question (2), namely ``yes but only if the
underlying logic contains $\exists\forall$ quantification''.

Of course, both these answers are only approximate and only speak in
broad terms. Whether each is to be answered with ``yes'' or ``no''
depends on the precise form of the logic, and on what kind of blow-up
we are willing to tolerate. To make matters concrete, we now consider
three particular logics, namely first-order predicate logic,
description logic (more specifically a particular version of $\ALC$)
and propositional logic (i.e., first-order predicate logic given a
finite universe and without quantification). We start with the latter.

In what follows, say we need to check whether formula $f$ is permitted
in $\DB$ under circumstances $\caseDesc$. We abstract from the
complications entailed by maintaining hierarchical consistency, and
assume that for $\crt$, it holds that $\mustref(\crt)=\emptyset$.

\begin{theorem}\label{theorem:permissibility:complexity:propositional}
For propositional logic, deciding permissibility is
$\Sigma^p_2$-complete.
\end{theorem}
{\it Proof sketch.} The set $\Sigma^p_2 = \NP^{\NP}$, so containment
is shown by guessing a supporting set and verifying its properties
using an $\NP$ oracle. For the hardness we encode an QBF formula
$\exists x\forall y: \phi(x,y)$ in permissibility request for case law
database. We do this by encoding all possible values for $x$ in the
database and asking for the permissibility of $\phi(x,y)$. Details can
be found in Appendix\cutShort{~\ref{app:perm:prop}}.
\vspace{1em}

As entailment testing in propositional logic is only $\coNP$-complete,
Theorem~\ref{theorem:permissibility:complexity:propositional} answers
question (1) with ``yes'', and answers question (2) with ``no, unless
we are willing to tolerate worst-case exponentially large
formulas''.\cutShort{ Unsurprisingly, the answers for first-order logic are
different:}

\begin{theorem}\label{theorem:permissibility:complexity:firstorder}
  Permissibility is equivalent to satisfiability of a formula whose
  size is polynomial in the size of $\DB$, $\caseDesc$, and $f$ for
  \vspace{-\topsep}\begin{enumerate}[(1)]
  \item first-order logic.
  \item the description logic $\ALC$ with concept constructors
    $\conFills$ and $\conOneOf$ by role constructors $\roleAnd$,
    $\roleNot$, $\roleProd$, and $\roleInverse$.\footnote{For details
      on this instance of $\ALC$, please consult
      \cite{borgida1996relative}.}
  \end{enumerate}
\end{theorem} {\it Proof sketch.} The result
in~\cite{borgida1996relative} shows equality of expressivity of
first-order logic with at most two free variables. Thus we construct a
suitable formula for the first part. We do this by using existantial
quantification in order to choose a warranted supporting set and then
design the formula such that it is satisfiable if and only if the
consistency properties of the case holds that can be constructed from
that supporting set (i.e., the case potentially output by
Algorithm~\ref{alg:permissibility}). All parts that are not choosen by
the existantial quantifier will be equivalent to $\top$. Details can
be found in Appendix\cutShort{~\ref{app:perm:fol}}.


\paragraph{Norm extraction} 
As seen in Section~\ref{sec:privCases}, privacy cases induce normative
rules. The format of rules gives the advantage that these are easy to
enforce and bridge the gap towards privacy policies. As shown by
Theorem~\ref{theorem:norm:extract} we extract a norm for every case in
the database. \cutShort{The assumption is that the case is consistent with
respect to an underlying consistent privacy case law database $\DB$.}
The algorithm is postponed to the appendix\cutShort{
(Algorithm~\ref{alg:norm:extr})}. It basically turns the proof of
Theorem~\ref{theorem:norm:extract} into an algorithm transforming the
logical formula of the case's $\facts$.

Let $f$ be the size of the biggest formula in the leaves of $C$ and
$n$ the number of nodes in $C$. Then the size of the norm can become
$\mathcal{O}(2^f\cdot n+\abs{\pre_C})$. The computation needs
operations linear in that size.\cutShort{ However, there is no need for any
operations to decide $\models$ in order to solve this reasoning task.}



\section{Logic Selection}\label{sec:automation}

%

For modeling purposes \cutShort{--- naturally modeling the background
  knowledge base, the detailed aspects characterizing a case
  description, and the reasoning applied in arguments ---} as well as
for computational purposes \cutShort{--- effectively realizing the
  desired reasoning tasks ---} the choice of logic is, of course, of
paramount importance. The only hard requirement (``must have'') that
the logic, $\cal L$, must meet is: \vspace{-\topsep}
\begin{itemize}
\item[(i)] \textbf{Sufficient expressivity} to tackle our framework and
  reasoning tasks. Precisely, the minimal requirement is for $\cal L$ to
  provide a language $\cal L_{F}$ for formulas, with reasoning support
  for tests of the form (a) $\bigwedge_{\phi \in \Phi} \models \bot$ and
  (b) $\bigwedge_{\phi \in \Phi} \models \psi$: These are the only tests
  our reasoning tasks demand from the underlying logic. If $\cal L_{F}$
  is closed under conjunction and contains $\bot$ (as will be the case
  in our logic of choice), the requirement simply becomes to be able to
  test whether $\phi \models \psi$. 
\end{itemize}
\noindent The soft requirements (``nice to have'') on the logic are:
\vspace{-\topsep}
\begin{itemize}
\item[(ii)] \textbf{Suitable for modeling real-world phenomena and
  knowledge}, ideally an established paradigm for such modeling tasks.
\item[(iii)] \textbf{Decidability, and as low complexity as possible}, of the
  relevant reasoning (e.g., satisfiability checks; cf.\ (i)).
\item[(iv)] \textbf{Effective tool support} established and available.
\end{itemize}
\vspace{-\topsep}

\noindent What we have just outlined is essentially a ``wanted
poster'' for \emph{description logic (DL)}
\cite{dlhandbook:2003}. This is a very well investigated family of
fragments of first-order logic\cutShort{ (several decades of research
  in AI and related areas)}, whose mission statement is to provide a
language for modeling real-world phenomena and knowledge (ii), while
retaining decidability and exploring the trade-off of expressivity
vs.\ complexity (iii). Effective tool support (iv) has been an active
area for two decades. Every DL provides a language to describe
``axioms'', and even the most restricted DLs \cutShort{(in particular,
  the DL-Lite family \cite{calvanese:etal:jar-07} which constitutes
  the ``lower extreme'' of the DL complexity scale)} make it possible
to answer queries about the truth of an axiom relative to a
conjunction of axioms, which is exactly the 
test we require.

To make things concrete, we briefly consider the description logic
\emph{attributive concept language with complements}, for short
$\ALC$, which was introduced in 1991
\cite{schmidt:smolka:ai-91},\footnote{For a comprehensive overview of
  current techniques and results regarding $\ALC$, see \cite{BaHS07a}.}
and is widely regarded as the canonical ``basic'' description logic
variant (most other DLs extend $\ALC$, in a variety of directions).
Description logic is a form of predicate logic that considers only
$1$-ary and $2$-ary predicates, referred to as \emph{concepts} and
\emph{roles}, respectively. Assuming a set $N_C$ of concept names and
a set $N_R$ of role names, DL makes it possible to construct
\emph{complex concepts}, which correspond to a particular subset of
predicate-logic formulas with exactly one free variable. For $\ALC$,
the set of complex concepts is the smallest set \replShort{such
  that}{s.t.}  \vspace{-.5\topsep}\begin{enumerate}
\item $\top,\bot$ and every concept name $A\in N_C$ are complex
  concepts, and
\item if $C$ and $D$ are complex concepts and $r\in N_R$, then
  $C\sqcap D$, $C\sqcup D$, $\lnot C$, $\forall r.C$, and $\exists
  r.C$ are complex concepts.
\end{enumerate}\vspace{-.5\topsep}
Here, $\sqcap$ denotes concept intersection (logical conjunction),
$\sqcup$ denotes concept union (logical disjunction), and $\lnot C$
denotes concept complement (logical negation). $\forall r.C$ collects
the set of all objects $x$ such that, whenever $x$ stands in relation
$r$ to $y$, $y \in C$. Similarly, $\exists r.C$ collects the set of
all objects $x$ such that there exists $y$ where $x$ stands in
relation $r$ to $y$ and $y \in C$.

$\ALC$ allows \emph{concept inclusion} axioms, of the form $C
\sqsubseteq D$, where $C,D$ are complex concepts, meaning that $C$ is
a subset of $D$ (universally quantified logical implication). $\ALC$
furthermore allows \emph{assertional} axioms, of the form $x : C$ or
$(x,y) : r$, where $C$ is a complex concept, $r$ is a role, and $x$
and $y$ are individual names (i.e., constants).  An $\ALC$ knowledge
base consists of finite sets of concept inclusion axioms and
assertional axioms (called the \emph{TBox} and \emph{ABox}
respectively), interpreted as conjunctions. The basic reasoning
services provided by $\ALC$ (and most other DLs) are testing whether a
knowledge base $\KB$ is satisfiable, and testing whether
$\KB\models\phi$ where $\phi$ is an axiom. These decision problems are
decidable, and more precisely, $\exptime$-complete for $\ALC$. (In
some DL-Lite variants, the decision problems are in \NP, or even
polynomial-time solvable.)

For our purposes, we can assume as our formulas $\cal L_{F}$
conjunctions of axioms, i.e., the smallest set that contains $\bot$,
all axioms of the underlying DL (e.g., $\ALC$), as well as $\phi
\wedge \psi$ if $\phi$ and $\psi$ are members of $\cal L_{F}$. In
order to test whether $\phi \models \psi$, we then simply call the DL
reasoning service ``$\phi \models \psi_i$?'' for every conjunct
$\psi_i$ of $\psi$ and return ``yes'' iff all these calls did. In
other words, we may use conjunctions of DL axioms in the knowledge
base, case descriptions, and proof tree nodes.

\section{Conclusion\lenSect{0,5}}\label{sec:conclusion}

In this paper, we introduced \fworkName, the first framework for
automated reasoning about case law. We showed that it complies with
natural requirements of consistency\cutShort{ and tailored the
  framework for privacy case law}. Moreover, we showed a tight
connection between privacy case law and the notion of norms that
underlies existing rule-based privacy research. We identified the
major reasoning tasks such as checking the case law database for
consistency, extracting norms and deducing whether an action is legal
or not. For all these tasks, we gave algorithms deciding them and we
did an analysis that leads to $\ALC$ as a suitable instantiation for
the logic. \cutShort{In particular, $\ALC$ provides efficient
  realizations while being sufficiently expressive and suitable for
  modeling real-world phenomena and knowledge.}

\cutShort{For future research, we need to construct a significantly
  large data base consisting of real world cases. Here, the challenge
  is to differentiate between statements made as world knowledge
  statement, those made because of the case descriptions and those
  referenced. The reason for this is that there is no clean
  language-wise separation in the argumentation.}


\paragraph{Acknowledgements}%
We want to thank the anonymous reviewer for their valuable
feedback. We tried to incorporate the feedback as much as
possible. \replShort{}{Due to space constraints, parts of the
  feedback was only used for the long version~\cite{BaBeHoMa:2015}.}

This work was supported by the German Ministry for Education and
Research (BMBF) through funding for the Center for IT-Security,
Privacy and Accountability (CISPA).


\appendix

\end{sloppypar}

 \bibliographystyle{abbrv}
 \bibliography{biblio}

\cutShort{
 
\section{Postponed Algorithms for Reasoning Tasks}\label{appendix:reasoning}

\subsection{Database Consistency}
Here, we present the algorithms for consistency that were postponed in
Section~\ref{sec:reason:task}. Algorithm~\ref{alg:case:consistent} can
be used to decide case consistency,
Algorihtm~\ref{alg:referential:consistent} can be used to decide
referential consistency and
Algorithm~\ref{alg:hierarchical:consistent} can be used to decide
hierarchical consistency.

\IncMargin{1em}
\RestyleAlgo{boxruled}
\LinesNumbered
\begin{algorithm}[htb!]
  \Indm
  \SetKwInOut{Input}{Input}
  \SetKwInOut{Output}{Output}
  \caption{Case consistency\label{alg:case:consistent}}
  \Input{A case $C = (\confPred,\caseDesc,\prTree,\crt)$}%
  \Output{$\top$ if $C$ is consistent and $\bot$ otherwise} %
  \Indp%
  Check that
  $\KB_W\wedge\caseDesc\models\pres_C$.\\
  Check that
  $\KB_W\wedge\caseDesc\wedge\facts_C\not\models\bot$.\\
  For every leaf node $n$ in $\prTree$ labeled with $\labAxiom$, check that $\KB_W\wedge\caseDesc\models n$.\\
  For every inner node $n$ in $\prTree$ annotated by $\AND$ with
  child nodes
  $n_1,\ldots,n_k$, check that $\bigwedge_{1\leq i\leq k}n_i\models n$.\\
  For every inner node $n$ in $\prTree$ annotated by $\OR$ with
  child nodes
  $n_1,\ldots,n_k$, check that $\bigvee_{1\leq i\leq k}n_i\models n$.\\
  If all checks succeed output $\top$; otherwise output $\bot$.
\end{algorithm}
 \DecMargin{1em}

\IncMargin{1em}
\RestyleAlgo{boxruled}
\LinesNumbered
\begin{algorithm}[htb!]
  \Indm
  \SetKwInOut{Input}{Input}
  \SetKwInOut{Output}{Output}
  \caption{Referential consistency\label{alg:referential:consistent}}
  \Input{A case $C = (\confPred,\caseDesc,\prTree,\crt)$ and a case law
    database $\DB$}%
  \Output{$\top$ if $C$ is referentially consistent w.r.t. $\DB$ and
    $\bot$ otherwise} %
  \Indp %
  \For{every subcase $D$ referenced by leaf node
    $\pre\rightarrow\fact$}{%
    check that $\KB_W\wedge\caseDesc\wedge\pre\models\pres_D$ } %
  If all checks succeed output $\top$; otherwise output $\bot$.
\end{algorithm}
\DecMargin{1em}

\IncMargin{1em}
\RestyleAlgo{boxruled}
\LinesNumbered
\begin{algorithm}[htb]
  \Indm
  \SetKwInOut{Input}{Input}
  \SetKwInOut{Output}{Output}
  \caption{Case-wise hierarchical consistency\label{alg:hierarchical:consistent}}
  \Input{A case $C = (\confPred,\caseDesc,\prTree,\crt)$ and a
    hierarchically consistent CLD $\DB$}%
  \Output{$\top$ if $\DB\cup\{C\}$ is hierarchically consistent (where
    $C$ is set to be the maximum w.r.t. $\leq_t$)} %
  \Indp%
  \For{every $D\in\DB$ for which $\crt\lecourt\crt_D$}{%
    check that $\KB_W\wedge\pres_C\wedge\pres_D\not\models\bot$\\
    check that
    $\KB_W\wedge\confPred\wedge\confPred_D\models\bot$.\\
    If both checks succeed output $\bot$.%
  }%
  Output $\top$.
\end{algorithm}
 \DecMargin{1em}

\subsection{Algorithm for Norm Extraction}

Algorithm~\ref{alg:norm:extr} can be used in order to extract a norm
from a privacy case.

\IncMargin{1em}
\RestyleAlgo{boxruled}
\LinesNumbered
\begin{algorithm}[htb!]
  \Indm
  \SetKwInOut{Input}{Input}
  \SetKwInOut{Output}{Output}
  \caption{Norm extraction\label{alg:norm:extr}}
  \Input{A case $C = (\confPred,\caseDesc,\prTree,\crt)$}%
  \Output{A norm $\phi$ that decides $\confPred$} %
  \Indp
  $phi := \top$\\
  \For{leaf node $n$ in $C$}{%
    $\phi := \phi \wedge$CNF$(n)$
  }%
  Remove $\lnot\confPred$ from $\phi$\\
  Remove all clauses not containing $\confPred$\\
  Remove $\confPred$ from $\phi$\\
  $\phi := \pre_C\wedge\lnot\phi$\\
  Output $\phi$
\end{algorithm}
\DecMargin{1em}


 \clearpage
 \section{Postponed proofs}\label{sec:proofs}


\subsection{Proof of Theorem~\ref{thm:deducSets}}

\begin{proof}
  We define $\caseDesc := A$ for a predicate $A$ and consider the
  court set $\courtSet = \{H^1_1,H^1_2,H^2\}$ such that $H^i\lecourt
  H^j$ iff $i < j$ implies $\mustref$ and $\mayref$ as in
  example~\ref{example:mustmay}.

  Let $\labDec(f)$ be a proof tree consisting of a single assessment
  node as root node that contains $\top \rightarrow f$ and, for a case
  $C$ and a formula $f$, let $\labCase(C,f)$ be the proof tree
  consisting of a single case reference node that refers to $C$ and
  contains the formula $\top \rightarrow f$. Let $B\neq A$ be some
  predicate. The database $\DB$ consists of the following cases:
\begin{itemize}
\item $C_1 = (p,\top,\labDec(p),H^1_1)$
\item $C_2 = (\lnot p,\top,\labDec(\lnot p),H^1_2)$
\item $C_3 = (A\Rightarrow B, \top, \labDec(A\Rightarrow B), H^1_1)$
\item $C_4 = (B\Rightarrow \lnot A, \top, \labDec(B\Rightarrow \lnot A), H^1_2)$
\end{itemize}

The time order $\leq_t$ is given by $<$ on the indices.

The database is obviously consistent. Let $\crt = H^2$.

\begin{enumerate}
\item\label{proof:perm:1} Define the set $F:=\{p,\lnot p\}$. The
  formula $p$ is permitted by $\DB$ for case description $\caseDesc$
  and court $\crt$, since $(p,\caseDesc,\labCase(C_1,p), \crt)$ is a
  case as required by Definition~\ref{def:deduc}. The same holds for
  $\lnot p$.

  Assume that $F$ is permitted. Then there are cases $C_p,C_{\lnot p}$
  such that $\DB\cup\{C_p,C_{\lnot p}\}$ is consistent. However, $C_p$
  and $C_{\lnot p}$ are in conflict and are at the same court level,
  i.e., either $\mustref(C_p,C_{\lnot p})$ holds or $\mustref(C_{\lnot
    p},C_p)$ depending on the order in which the cases are inserted in
  $\DB$. As a consequence, $\DB\cup\{C_p,C_{\lnot p}\}$ cannot be
  hierarchically consistent. Thus, that database cannot be consistent
  either. Therefore, $F$ cannot be permitted.

\item Let $f_1 = A\Rightarrow B$, $f_2 = B\Rightarrow \lnot A$, and
  $F=\{f_1,f_2\}$. It is easy to see that for a case $C_{f_1\wedge
    f_2}$ it holds that $\KB_W\wedge\caseDesc\wedge\facts_C\models\bot$
  if $C_3$ and $C_4$ are referenced. That means the case is not
  consistent. However, without referencing these cases it is
  impossible to prove $f_1\wedge f_2$ as a decision formula within
  $\DB$.

  The set $F$ is permitted. Since $C_{f_1},C_{f_2}$ as constructed in
  the proof of~\ref{proof:perm:1} are consistent. These cases are also
  not in conflict. In order to prove the absence of a conflict, we have to
  check that $\KB_W\wedge\pres_{C_1}\wedge\pres_{C_2}\not\models\bot$
  and $\KB_W\wedge\facts_{C_1}\wedge\facts_{C_2}\models\bot$. While
  the first condition is met, the second does not hold, since we need
  $\caseDesc=A$ to entail $\bot$.
\end{enumerate}
\end{proof}


\subsection{Proof of Theorem~\ref{theorem:allFollowsFromDecisions}}

\begin{proof}
  We prove the theorem step by step in the same order the claims are
  defined.

\begin{enumerate}
\item We show a stronger statement for $\caseDesc=\pres_C$ (since $C$ is
consistent it has to hold that $\KB_W\wedge\caseDesc\models\pres_C$).

We start with $\mathcal{A}^C$ as the set of all leaf nodes of $C$ that
are annotated by $\labDec$ and $\labCase(D)$ for some $D$. For this
set all properties (\ref{def:sup:set:p1})--(\ref{def:sup:set:p3}) of
Defininition~\ref{def:sup:set} clearly hold by consistency of
$C$. However, the set might contain nodes labeled with $\labCase(D)$
which we need to replace in order to fulfill this criterion of the
Theorem, as well.

For a fixed leaf formula $(\pre\rightarrow\fact)\in\mathcal{A}^C$
corresponding to a $\labCase(D)$ leaf node, take the set $\mathcal{A}^D$
for $D$ defined as $\mathcal{A}^C$ for $C$. By consistency of $D$, we
get (a) and (b) for $\caseDesc_D=\pres_D$ and $f=\fact$. By referential
consistency it holds that $\KB_W\wedge\pre\models\pres_D$. Therefore, if
we replace $\mathcal{A}^C$ by
$\mathcal{A}^C\backslash\{(\pre\rightarrow\fact)\}\cup\mathcal{A}^D$,
property (a) holds for $C$ and the new $\mathcal{A}^C$ since
$\KB_W\wedge\pre\models\pres_D$. Property (b) also transfers to the new
set, since (b) holds for the old set and (b) holds for $D$ and
$\mathcal{A}^D$ with respect to $\caseDesc_D=\pres_D$ and $f=\fact$.

The process of successively replacing $\labCase(D)$ nodes in
$\mathcal{A}^C$ terminates since $\mathcal{A}^D$ only contains
$\labCase(E)$ leaf nodes for $E\leref D$ and $\DB$ is finite.

Our proof above actually shows that
$\KB_W\wedge\pres_C\wedge\bigwedge_{(\pre\rightarrow\fact)\in\mathcal{A}}
\fact\models \facts_C$, hence (c) follows from consistency of $C$.

\item The direction $\Rightarrow$ follows from the first part of the
  proof since permissibility implies that we can add a case as
  specified. So consider $\Leftarrow$, i.e., let $\mathcal{A}$ be a
  set supporting for $f$ in circumstances $\caseDesc$ for a court
  $\crt$.

  We can construct a case $C$ by referencing all these decisions and
  putting $f$ in a root node that has all these references as
  child nodes. The properties
  (\ref{def:sup:set:p1})--(\ref{def:sup:set:p3}) of $\mathcal{A}$
  (Definition~\ref{def:sup:set}) imply consistency of $C$. The
  requirement that the nodes are warranted and that $C$ is at
  the end of the timeline implies that we reference correctly.

  The $\DB\cup\{C\}$ is also hierarchically consistent since $C$ does
  not introduce new conflicts. Otherwise $\mathcal{A}$ would already
  be in conflict with $\DB$.

\item The direction $\Rightarrow$ follows immediately from the
  previous part of the proof since $f$ is deducible if $f$ is
  permitted and $\lnot f$ is not permitted. The other direction also
  follows from the previous part since the existence of $\mathcal{A}$
  implies that $f$ is permitted and the non-existence of support for
  $\lnot f$ is implied by the requirement of $\mathcal{B}$.
\end{enumerate}

\end{proof}


\subsection{Proof of Theorem~\ref{theorem:consistencyNecessary}}

\begin{proof} 
  Let $\crt$ be a court with $\mustref(\crt)=\emptyset$. For $C_1$,
  select an arbitrary $D\in\DB$, and construct $\prTree$ containing
  root node $f$ and a single leaf node $(\top\rightarrow f)$ labeled
  with $\labCase(D)$. Define $C_1 := (f,\top,\prTree,\crt)$. Then
  $\DB\cup\{C_1\}$ is case-wise consistent since $\DB$ is case-wise
  consistent (note that we do not enforce referential consistency, so
  ignore whether or not $f$ is actually decided by $D$). Hierarchical
  consistency holds simply because $C_1$ does not need to reference
  other cases.

  For $C_2$,construct $\prTree$ containing the single node $f$ labeled
  with $\labAxiom$. Define $C_2 := (f,\top,\prTree,\crt)$. This case is
  not consistent; however, $\DB\cup\{C_2\}$ is referentially consistent
  simply because $C_2$ does not make any references. Hierarchical
  consistency holds for the same reason as before.
\end{proof}

\subsection{Proof of Theorem~\ref{theorem:botBringsConflicts}}

\begin{proof}
  For $\bot$, this holds simply because deducibility requires us to
  construct a consistent case with root node $\bot$, and any case $C$
  one of whose nodes is $\bot$ is not consistent. To see the latter,
  just note that, if $C$ was consistent, then by
  Definition~\ref{def:case:consist} (v) it follows that
  $\facts_C\models\bot$, which by Definition~\ref{def:case:consist}
  (iii) means that $C$ is not consistent.

  For $\{f, \neg f\}$, assume to the contrary that there exist cases
  $C_f=(f,\caseDesc,$ $\prTree_f,\crt)$ and $C_{\neg f}=(\neg
  f,\caseDesc,\prTree_{\neg f},\crt)$ such that $\prTree_f$ and
  $\prTree_{\neg f}$ do not contain nodes labeled with $\labDec$, and
  $\DB\cup\{C_f, C_{\neg f}\}$ is consistent (where the new cases are
  inserted in any order at the end of the timeline $\leq_t$). But
  since $\crt\in\mustref(\crt)$, the latter one has to respect the
  first one. We show that $C_f$ and $C_{\neg f}$ are in conflict, thus
  contradicting the hierarchical consistency of $\DB\cup\{C_f, C_{\neg
    f}\}$. Obviously, $\KB_W\wedge\facts_{C_f}\wedge\facts_{C_{\neg
      f}}\models f\wedge\lnot f\models\bot$. It remains to show that
  $\KB_W\wedge\pres_{C_f}\wedge\pres_{C_{\neg f}}\not\models\bot$. By
  consistency of each of $C_f$ and $C_{\neg f}$, we get (a)
  $\KB_W\wedge\caseDesc\not\models\bot$, (b)
  $\KB_W\wedge\caseDesc\models\pres_{C_f}$ and (c)
  $\KB_W\wedge\caseDesc\models\pres_{C_{\neg f}}$. Putting (b) and (c)
  together gives
  $\KB_W\wedge\caseDesc\models\pres_{C_f}\wedge\pres_{C_{\neg f}}$,
  which with (a) shows
  $\KB_W\wedge\caseDesc\wedge\pres_{C_f}\wedge\pres_{C_{\neg
      f}}\not\models\bot$, which is stronger than what we needed to
  prove. Therefore, $\{f,\lnot f\}$ is not permitted in $\DB$.
\end{proof}


\subsection{Proof of Theorem~\ref{theorem:norm:extract}}

\begin{proof}
  We show the statement for $\confPred=\legalPred(a)$ for some
  $a$. The proof for $\lnot\legalPred(a)$ is analogous.  Given
  consistency of $C$, we get that $\facts_C\models\confPred$.
  Transforming $\facts_C$ to a CNF formula, we can write $\facts_C$ as
  $\phi_W\wedge\phi_L$ where $\legalPred(a)$ only occurs in
  $\phi_L$. Since $\phi_W\wedge\phi_L\models\legalPred(a)$ we can
  assume that $\phi_L$ does not contain
  $\lnot\legalPred(a)$. Otherwise we could remove the
  $\lnot\legalPred(a)$ maintaining the property of
  $\phi_W\wedge\phi_L\models\legalPred(a)$.

  Every literal $l_j$ of the formula $\phi_L$ has the form
  $\legalPred(a)\vee\bigvee_{1\leq i\leq k}x_i$, which is equivalent to
  $\underbrace{(\bigwedge_{1\leq i\leq k}\lnot
  x_i)}_{=: r_j}\Rightarrow\legalPred(a)$. Hence, we can write $\phi_L$ as
  $(\bigvee_{1\leq j\leq m} r_j)\Rightarrow\legalPred(a)$. We define
  $\phi_S := \bigvee_{1\leq j\leq m} r_j$ and
  get $$\phi_W\wedge(\phi_S\Rightarrow\legalPred(a))\models\legalPred(a)$$
  where neither $\phi_W$ nor $\phi_S$ contain
  $\legalPred(a)$. Therefore, it must hold that
  $\phi_W\models\phi_S$. However, this argumentation was only
  applicable in the case $C$ since
  $\KB_W\wedge\caseDesc\models\pres_C$. Hence we can derive the norm
  $\phi^+:=\pres_C\wedge\phi_S$ as positive norm.
\end{proof}


\subsection{Proof of Corollary~\ref{cor:norm:forms}}

\begin{proof}
The consistency of $N(C)$ follows from the previous theorem. The leaves
of $N(C)$ are the same as the leaves of $C$, and thus referentially
consistency follows from $C$'s referential consistency. In addition,
$\confPred$ of $N(C)$, as well as $\pres_C$ of $N(C)$, are the same as
of $C$, and thus $N(C)$ is in conflict with a case iff $C$ is. Therefore,
hierarchical consistency is also maintained.
\end{proof}


\subsection{Proof of Theorem~\ref{theorem:permissibility:complexity:propositional}}\label{app:perm:prop}

\begin{proof}
  Recall that $\Sigma^p_2 = \NP^{\NP}$. Membership follows because we
  can guess the set $\mathcal{A}$ and check, using an $\NP$ oracle,
  the three entailment tests
  (\ref{def:sup:set:p1}--\ref{def:sup:set:p3}). The consistency of the
  set with $\DB$ can also be answered by the $\NP$ oracle since
  verifying a conflict can be done in polynomial time.

  For hardness, consider a QBF formula of the form $\exists X \forall
  Y \phi(X,Y)$ where each of $X$ and $Y$ are variable sets and
  $\phi(X,Y)$ is an arbitrary propositional formula in the variables
  $X \cup Y$. Testing validity of $\exists X \forall Y \phi(X,Y)$ is
  $\Sigma^p_2$-hard. To polynomially reduce this to permissibility
  testing over a propositional logic, we construct a corresponding
  case law database $\DB$ as follows. For each $x \in X$, $\DB$
  includes a case $(x,\top,\prTree,\crt)$ where $\prTree$ consists of
  a single $\labDec$ node of the form $\top \rightarrow x$, as well as
  a case $(\neg x,\top,\prTree,\crt)$ where $\prTree$ consists of a
  single $\labDec$ node of the form $\top \rightarrow \neg x$. In
  other words, for each $x$ we have both truth-value decisions
  available for $\mathcal{A}$ to choose from. We set $f :=
  \phi(X,Y)$. Obviously, this reduction is polynomial in the size of
  the formula $\exists X \forall Y \phi(X,Y)$. To see that the
  reduction is correct, observe that $f$ is permitted in $\DB$ iff
  there exists a truth assignment $a$ to $X$ which, viewed as a
  conjunction of literals, entails $\phi(X,Y)$, i.e.,\ $a \models
  \phi(X,Y)$. The latter is the case iff there exists $a$ s.t., for
  all truth assignments to $Y$, $\phi(a(X),Y)$ is true (where
  $\phi(a(X),Y)$ instantiates each $x \in X$ with $a(x)$). This,
  finally, is the case iff $\exists X \forall Y \phi(X,Y)$ is valid,
  which is what we needed to show.
\end{proof}


\subsection{Proof of Theorem~\ref{theorem:permissibility:complexity:firstorder}}\label{app:perm:fol}

\begin{proof}
  According to~\cite{borgida1996relative}, the expressiveness of the
  description logic $\ALC$ extended concept constructors $\conFills$
  and $\conOneOf$ by role constructors $\roleAnd$, $\roleNot$,
  $\roleProd,$ and $\roleInverse$ is equal to the expressiveness of
  first-order predicate logic with predicates of arity at most $2$ and
  at most $2$ free variables (in any subformula). Consequently, we
  show that the construction for first-order logic increase neither
  the arity of predicates nor the number of free variables.

  Let $\mathcal{L} = \{n_1=(\pre_1\rightarrow\fact_1), \dots,
  n_k=(\pre_k\rightarrow\fact_k)\}$ be the set of all warranted leaf
  formulas of cases $C'\in\DB$ with label $\labDec$. We need to
  construct a first-order formula $\phi$ that is valid iff there
  exists $\mathcal{A} \subseteq \mathcal{L}$ such that the three
  implications (1--3) of Definition~\ref{def:sup:set} hold. Our idea
  is to encode the choice of that subset as an ``on/off switch''
  associated with each $n_i$. The switch will be realized through an
  existential quantifier over $x_1,\ldots,x_k$ and a unary predicate
  $\chosen_i$ for every $i\in\{1,\ldots,k\}$ which we add to the FOL
  signature (w.l.o.g. all $\chosen_i$ do not occur in any $\pre_i$ or
  $\fact_i$).  The meaning of the predicate is that $\chosen_i(x_i)$
  holds if and only if $n_i$ is chosen for the set $\mathcal{A}$.

  We next define the formulas $\phipre_i := (\neg \chosen_i(x_i) \vee
  \pre_i)$ and $\phifact_i := (\neg \chosen_i(x_i) \vee \fact_i)$ to
  implement our switches. Note that, if for $x_i$ it holds that
  $\lnot\chosen_i(x_i)$, then both $\phipre_i$ and $\phifact_i$
  simplify to $\top$; if for $x_i$ it holds that $\chosen_i(x_i)$,
  then $\phipre_i$ simplifies to $\pre_i$ and $\phifact_i$ simplifies
  to $\fact_i$. Using these building blocks, we define our
  correspondences to the implications (1--3), as follows:
  \begin{itemize}
  \item[(1)] $\phia$ := $\KB_W \wedge \caseDesc \Rightarrow
    \bigwedge_{i=1}^k \phipre_i$.
  \item[(2)] $\phib$ := $\KB_W \wedge \caseDesc \wedge
    \bigwedge_{i=1}^k \phifact_i \Rightarrow f$.
  \item[(3)] $\phic$ := $\neg (\KB_W \wedge \caseDesc \wedge
    \bigwedge_{i=1}^k \phifact_i \Rightarrow \bot)$.
  \end{itemize}
  Our formula $\phi$ then is defined simply as $\phi := \phia \wedge
  \phib \wedge \phic$. We now prove that $\phi$ is satisfiable iff
  there exists $\mathcal{A} \subseteq \mathcal{L}$ such that the three
  implications (a--c) hold.

  ``$\Leftarrow$'': Assume there is a set $\mathcal{A}$ such that the
  implications (1--3) hold. We define an assignment $a$ for the $x_i$
  as follows: if $(\pre_i\rightarrow\fact_i)\in\mathcal{A}$, then
  $\chosen_i(x)\equiv\top$ and otherwise
  $\chosen_i(x)\equiv\bot$. Then $\bigwedge_{i=1}^k\phipre_i$ reduces
  to $\bigwedge_{(\pre\rightarrow\fact)\in\mathcal{A}}\pre$ and
  $\bigwedge_{i=1}^k \phifact_i$ reduces to
  $\bigwedge_{(\pre\rightarrow\fact)\in\mathcal{A}}\fact$. Thus, (1)
  implies $I,a\models\phia$, (2) implies $I,a\models\phib$ and (3)
  implies $I,a\models\phic$ for every FOL interpretation
  $I$. Consequently, $\phi$ is satisfiable.

  ``$\Rightarrow$'': Now assume $\phi$ is satisfiable, i.e., there is
  an interpretation $I$ such that $I \models\phi$ holds. Therefore,
  there is an assignment $a$ for the $x_i$ such that
  $I,a\models\phia\wedge\phib\wedge\phic$. For such an assigment $a$,
  we define $\mathcal{A}:=\{(\pre_i\rightarrow\fact_i)\mid
  I,a\models\chosen_i(x_i)\}$.  For this set $\mathcal{A}$ the
  formulas $\phia,\phib,\phic$ can be reduced as in ``$\Leftarrow$'',
  i.e., the conditions (1--3) hold.

  We cannot apply the result of~\cite{borgida1996relative} directly,
  since the $x_1,\ldots,x_k$ introduce many free variables in
  $\phia\wedge\phib\wedge\phic$. To clarify how we can reduce this
  number, we consider the formula $\exists x_1,\ldots,x_k: \phi$ which
  is satisfiable iff $\phi$ is satisfiable.

  The formula $\phic$ is logically equivalent to
  $\KB_W\wedge\caseDesc\wedge\bigwedge_{i=1}^{k}\phifact_i$. Calling
  this formula $\psic$, it follows that $\phib\wedge\phic$ is
  equivalent to $\psic\wedge f$ and $\phia\wedge\phib\wedge\phic$ is
  equivalent to $\psic\wedge f\wedge \bigwedge_{i=1}^{k}\phipre_i$. By
  reordering the conjunctive literals, we get
  \[ \KB_W\wedge\caseDesc\wedge f \wedge\bigwedge_{i=1}^{k}
  (\phipre_i\wedge\phifact_i) \]

  By definition $\phipre_i\wedge\phifact_i$ is equivalent to
  $\psi_i(x_i) = \lnot\chosen_i(x_i)\vee(\pre_i\wedge\fact_i))$. Now,
  the variable $x_i$ occurs only once in the whole formula. This
  allows us to rewrite $\phi$ as formula $\psi :=$
  \[ \KB_W\wedge\caseDesc\wedge f\wedge\bigwedge_{i=1}^k ((\exists
  x_i:\lnot\chosen_i(x_i)) \vee (\pre_i\wedge\fact_i)) \] Here it is
  easy to see that the transformations of~\cite{borgida1996relative}
  are applicable to the formula $\psi$ leading to a description logic
  expression if and only if they are applicable to $\KB_W, \caseDesc,
  f, \pre_i,$ and $\fact_i$. However, since we these formulas are
  formulated in the same description logic, it follows that the
  mentioned transformation is applicable leading to a description
  logic expression for $\psi$.
\end{proof}



}

\end{document}